\documentclass[amssymbols,12pt]{article}
  \usepackage{amssymb}
  \usepackage{graphicx}
  \usepackage{psfrag}

\newcommand{\curl}[4]
{\left( \begin{array}{cc} #1 & #2 \\[2ex] #3 & #4 \end{array}
\right)}

\newcommand\A{ {\mathcal{A}} }
\newcommand\K{ {\mathcal{K}} }
\newcommand\diag{ {\mathrm{diag}} }

\newcommand\B{ {\mathcal{B}} }

\newcommand\LL{ {\mathcal{L}} }
\newcommand\MM{ {\mathcal{M}} }
\newcommand\be{\begin{eqnarray}}
\newcommand\barray{$$\begin{array}{rl}}
\newcommand\ee{\end{eqnarray}}
\newcommand\earray{\end{array}$$}

\newcommand\half{\frac{1}{2}}

\newcommand\QQ{\mathbb{Q}}

\newcommand\s{Schr\"odinger }

\begin{document}
\title{The Nonlinear \s Equation on the Interval}
\author{A.S. Fokas  \\
{\em Department of Applied Mathematics and Theoretical Physics } \\
{\em University of Cambridge } \\
{\em Cambridge, CB3 0WA, UK} \\
{\em t.fokas@damtp.ac.uk} \\
and \\
A.R. Its \\
{\em Department of Mathematical Sciences, IUPUI} \\
{\em Indianapolis, IN  46202-3216, USA} \\
{\em itsa@math.iupui.edu}}

\date{}

\maketitle

\vskip .2in

\begin{abstract}
Let $q(x,t)$ satisfy the Dirichlet initial-boundary value problem for the
nonlinear Schr\"odinger equation on the finite interval, $0 < x < L$, with
$q_{0}(x) = q(x,0)$, $g_{0}(t) = q(0,t)$, $f_{0}(t) = q(L,t)$. Let $g_{1}(t)$
and $f_{1}(t)$ denote the {\it unknown} boundary values $q_{x}(0,t)$ and
$q_{x}(L,t)$, respectively. We first show that these unknown functions can be
expressed in terms of the given initial and boundary conditions through
the solution
of a system of nonlinear ODEs. Although the question of the global
existence of
solution of this system remains open, it appears that this is the first
time in the literature that such a characterization is explicitely described
 for a nonlinear evolution PDE defined on the interval; this result is the extension of the analogous result of [4]
and [6] from the half-line to the interval. We then show that
$q(x,t)$ can be expressed in terms of the solution of a $2\times 2$ matrix
Riemann-Hilbert problem formulated in the complex $k$ - plane. This
problem has explicit $(x,t)$ dependence in the form $\exp[2ikx + 4ik^2t]$,
and it has jumps across the real and imaginary axes. The relevant jump
matrices
are explicitely given in terms of the spectral data $\{a(k), b(k)\}$,
$\{A(k), B(k)\}$, and $\{\A(k), \B(k)\}$, which in turn are defined in
terms of
$q_{0}(x)$, $\{g_{0}(t), g_{1}(t)\}$, and $\{f_{0}(t), f_{1}(t)\}$,
respectively.

\end{abstract}

\section{Introduction}

We analyse the Dirichlet initial-boundary value problem for the nonlinear
\s (NLS) equation on a finite interval:
$$ iq_t + q_{xx} - 2\lambda|q|^2q = 0, \quad \lambda = \pm 1, \quad 0<x<L,
\quad 0<t<T,$$
$$ q(x,0) = q_0(x), \quad 0<x<L,$$
$$ q(0,t) = g_0(t), \quad q(L,t) = f_0(t), \quad 0<t<T, \eqno (1.1)$$
where $L,T$ are positive constants, and $q_0$, $g_0$, $f_0$, are smooth
functions
compatible at $x=t=0$ and at $x=L$, $t=0$, i.e. $q_0(0) = g_0(0)$, $q_0(L)
= f_0(0)$.

Our analysis is based on the extension of the results of [4] and [6] from the half-line
to the interval.

The analysis involves three steps.

\paragraph{ Step 1: A RH formulation under the assumption of existence.} \ \

We {\it assume} that there exists a smooth solution $q(x,t)$.

We use the {\it simultaneous} spectral analysis of the associated Lax pair
of the NLS
to express $q(x,t)$ in terms of the solution of a $2\times 2$-matrix
Riemann-Hilbert
(RH) problem defined in the complex $k$-plane.  This problem has {\it
explicit} $(x,t)$
dependence in the form of $\exp\{ 2i(kx+2k^2t)\}$, and it is uniquely
defined in terms
of the so-called {\it spectral functions},
$$ \{ a(k),b(k) \}, \quad \{ A(k),B(k) \}, \quad \{ \A (k), \B (k) \}. \eqno
(1.2)$$
These functions are defined in terms of
$$ q_0(x), \quad \{ g_0(t), g_1(t) \}, \quad \{ f_0(t), f_1(t) \}, \eqno
(1.3)$$
respectively, where $g_1(t)$ and $f_1(t)$ denote the {\it unknown}
boundary values
$q_x(0,t)$ and $q_x(L,t)$.

We show that the spectral functions (1.2) are {\it not} independent but
they satisfy
the {\it global relation}
$$ \left(a\A + \lambda \bar b e^{2ikL} \B \right) B- \left( b\A + \bar a
e^{2ikL} \B
\right) A = e^{4ik^2T} c^+(k), \quad k \in {\mathbb{C}}, \eqno (1.4)$$
where $c^+(k)$ is an entire function which is of $O(1/k)$ as $k
\rightarrow \infty$,
Im$k >0$; in fact,
$$
c^{+}(k) = O\left(\frac{1+ e^{2ikL}}{k}\right), \quad k \to \infty.
$$

\paragraph{Step 2: Existence under the assumption that the spectral
functions satisfy
the global relation.} \ \

Motivated from the results of Step 1, we {\it define} the spectral
functions (1.2) in
terms of the smooth functions (1.3).  We also {\it define} $q(x,t)$ in
terms of the
solution of the RH problem formulated in Step 1.  We {\it assume} that
there exist
smooth functions $g_1(t)$ and $f_1(t)$ such that the spectral functions
(1.2) satisfy
the global relation (1.4).  We then prove that: (i) $q(x,t)$ is defined
globally for
all $0<x<L$, $0<t<T$.  (ii) $q(x,t)$ solves the NLS equation.  (iii)
$q(x,t)$ satisfies
the given initial and boundary conditions, i.e., $q(x,0) = q_0(x)$,
$q(0,t) = g_0(t)$,
$q(L,t) = f_0(t)$.   A byproduct of this proof is that $q_x(0,t) = g_1(t)$
and
$q_x(L,t) = f_1(t)$.

\paragraph{Step 3: The analysis of the global relation.}  \ \

Given $q_0$, $g_0$, $f_0$, we show that the global relation (1.4)
characterizes $g_1$
and $f_1$ through the solution of a system of nonlinear Volterra integral
equations.
The rigorous investigation of these equations remains open.

We now discuss further the above three steps.

The analysis of {\it Step 1} is based on the introduction of appropriate
eigenfunctions
which satisfy {\it both} parts of the Lax pair.  It was shown in \cite{1} that for
linear PDEs defined in a polygonal domain with $N$ corners,
there exists a canonical way of
choosing such eigenfunctions: There exist $N$ such eigenfunctions each  of
them normalized
with respect to each corner.  Motivated by this result we introduce four
eigenfunctions,
$\{
\mu_j(x,t,k) \}^4_1$, see Fig.~\ref{fig1.1},
such that

$$ \mu_1(0,T,k) = I, \quad \mu_2(0,0,k) = I, \quad \mu_3(L,0,k) = I, \quad
\mu_4(L,T,k)
= I, \eqno (1.5)$$
where  $\mu_j$ are $2\times 2$ matrices and $I =$ diag$(1,1)$.  It can be
shown that
these
\begin{figure}[h]
\psfrag{1}{1}
\psfrag{2}{2}
\psfrag{3}{3}
\psfrag{4}{4}
\psfrag{y}{$y$}
\psfrag{t}{$\tau$}
\psfrag{T}{$T$}
\psfrag{x}{$(x,t)$}
\begin{center}
\includegraphics{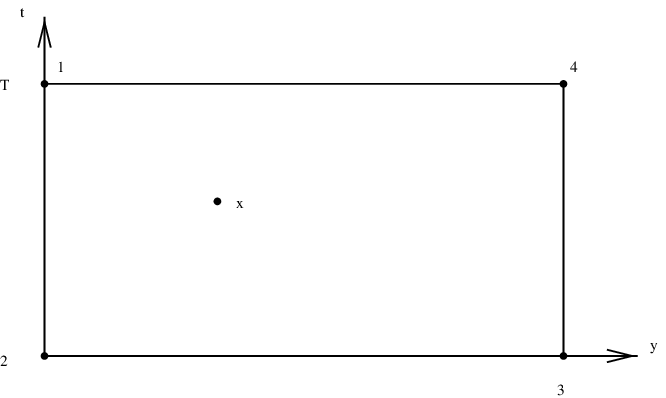}
\end{center}
\end{figure}\label{fig1.1}

\centerline{\bf Fig. 1.1}
\noindent eigenfunctions are simply related through the three matrices
$s$, $S$,
$S_L$,
$$ s(k) = \mu_3(0,0,k), S(k) = \left( e^{2ik^2T\sigma_3}
\mu_2(0,T,k)e^{-2ik^2T\sigma_3}\right)^{-1}, S_L(k) = \left(
e^{2ik^2T\sigma_3}
\mu_3(L,T,k)e^{-2ik^2T\sigma_3}\right)^{-1}, \eqno (1.6)$$
where $\sigma_3 = $diag$(1,-1)$.  These matrices satisfy certain symmetry
properties,
thus they can be represented by
$$ s(k) = \curl{\overline{a(\bar k)}}{b(k)}{\overline{\lambda b(\bar
k)}}{a(k)}, \quad
S(k) = \curl{\overline{A(\bar k)}}{B(k)}{\overline{\lambda B(\bar
k)}}{A(k)}, \quad
S_L(k) = \curl{\overline{\A(\bar k)}}{\B(k)}{\overline{\lambda \B(\bar
k)}}{\A(k)}. \eqno (1.7)$$

Regarding {\it Step 2} we note that equations (1.6), (1.7) motivate the
following definitions: Let the vectors
$$ (\phi_1(x,k),\phi_2(x,k))^\dag, \quad (\Phi_1(t,k),\Phi_2(t,k))^\dag,
\quad
(\varphi_1(t,k),\varphi_2(t,k))^\dag, \eqno (1.8a)$$
solve the $x$-part of the Lax pair evaluated at $t=0$, the $t$-part of the
Lax
pair evaluated at $x=0$, and the $t$-part of the Lax pair evaluated at $x=L$,
respectively, and let these vectors satisfy the boundary conditions
$$(\phi_1(L,k),\phi_2(L,k))^\dag = (0,1)^\dag, \quad
(\Phi_1(0,k),\Phi_2(0,k))^\dag =
(0,1)^\dag,\quad (\varphi_1(0,k),\varphi_2(0,k))^\dag = (0,1)^\dag. \eqno
(1.8b)$$
Define the spectral functions (1.2) by
$$ a(k)=\phi_2(0,k), \quad b(k) =\phi_1(0,k), $$
$$
A(k) = A(T,k), \quad B(k) = B(T,k), \quad \A(k) = \A(T,k),
\quad \B(k) = \B(T,k),
$$
where
$$A(t,k) = \overline{\Phi_2(t,\bar k)},\quad B(t,k) =-e^{4ik^2t}\Phi_1(t,k),
\eqno (1.9) $$
$$\A(t,k) = \overline{\varphi_2(t,\bar k)},\quad \B(t,k) = -e^{4ik^2t}
\varphi_1(t,k). $$

We note that the functions (1.2) depend on the functions (1.3).

The global existence of $q(x,t)$ is based on the unique solvability of the
associated
RH problem, which in turn is based on the distinctive nature of the
functions defining
the jump matrices: these functions have explicit $(x,t)$ dependence in an
exponential
form and they involve the spectral functions $s(k)$, $S(k)$, $S_L(k)$,
which have the
symmetry properties expressed in equations (1.7).  Using these facts it
can be shown
that the associated homogeneous RH problem has only the trivial solution
(i.e. there
exists a vanishing lemma).  The proof that $q(x,t)$ solves the given
nonlinear PDE uses
the standard arguments of the dressing method \cite{2}.  The proof that $q(0,t)
= q_0(x)$, is
based on the fact that the RH problem satisfied at $t=0$ is equivalent to
a RH problem
defined in terms of $s(k)$ which characterizes $q_0(x)$. The proofs that $\{
\partial^l_x q(0,t) = g_l(t)\}^1_0$ and that $\{ \partial^l_x q(L,t) =
f_l(t)\}^1_0$,
make crucial use of the global relation (1.4).  Indeed, it can be shown
that the RH
problems at $x=0$ and at $x=L$, are equivalent to RH problems involving
only $S(k)$ and
$S_L(k)$ (which in turn characterize $\{ g_l(t)\}^1_{l=0}$ and $\{
f_l(t)\}^1_{l=0}$, if and
only if the spectral functions satisfy the global relation.  Thus this
relation is not
only a necessary but it is also a {\em sufficient} condition for
existence.  Hence given $\{
q_0,g_0,f_0\}$, the main problem becomes to show that the global relation
characterizes
$g_1$ and $f_1$.

The analysis of {\it Step 3} is based on the Gelfand-Levitan-Marchenko
representation
of the eigenfunctions $\Phi=(\Phi_1,\Phi_2)^\dag$ and
$\varphi=(\varphi_1,\varphi_2)^\dag$.  Using these representation it can
be shown \cite{3}
that $\Phi$ can be expressed in terms of four functions $\{ M_j(t,s),
L_j(t,s)\}^2_1$,
$-t<s<t$, $t>0$, satisfying four PDEs (see \cite{3}) as well as the boundary
conditions
$$ L_1(t,t) = \frac{i}{2} g_1(t), \quad L_2(t,-t) = 0, \quad M_1(t,t) =
g_0(t), \quad
M_2(t,-t) =0. \eqno (1.10)$$
Similarly $\varphi$ can be expressed in terms of the four functions $\{
\MM_j(t,s),
\LL_j(t,s) \}^2_1$ satisfying
$$ \LL_1(t,t) = \frac{i}{2} f_1(t), \quad \LL_2(t,-t)=0, \quad \MM_1(t,t)
= f_0(t), \quad
\MM_2(t,-t) =0. \eqno (1.11)$$
Using the definitions (1.9) it can be shown that \cite{4}
$$A(t,k) = 1+ \int^t_0 e^{4ik^2\tau} \left[ 2\overline{L_2}(t,t-2\tau) -
i\lambda
g_0(t)\overline{M_1}(t,t-2\tau) + 2k\overline{M_2} (t,t-2\tau)\right]
d\tau,$$
$$ B(t,k) =- \int^t_0 e^{4ik^2\tau} \left[ 2L_1(t,2\tau-t)
-ig_0(t)M_2(t,2\tau-t) +
2kM_1(t,2\tau-t)\right]d\tau. \eqno (1.12)$$
Similar expressions are valid for $\A$ and $\B$ , where $L_1$, $M_1$,
$M_2$, $g_0$ are
replaced by $\LL_1$, $\MM_1$, $\MM_2$, $f_0$ respectively.
Substituting the expressions for $A,B,\A,\B$ in the global relation
(1.4) and letting $k \rightarrow -k$ in the resulting equation, we
obtain two relations coupling
$$g_0, f_0, L_1, M_1, M_2, \LL_1, \MM_1, \MM_2. \eqno (1.13)$$
It is remarkable that these two relations can be explicitly solved for
$g_1$ and $f_1$
in terms of the quantities appearing in (1.13).

Having solved the global relation it is now more convenient to formulate
the final
result in terms of the functions
$$\left\{ \hat L_j(t,k), \quad \hat M_j(t,k), \quad \hat{\LL}_j(t,k), \quad
\hat{\MM}_j(t,k) \right\}^2_{j=1}, \eqno (1.14)$$
where
$$\hat L_j(t,k) = \int^t_{-t} e^{2ik^2(s-t)} L_j(t,s)ds, \eqno (1.15)$$
and similarly for $\hat M_j$, $\hat{\LL}_j$, $\hat{\MM}_j$. Using these
notations,
the explicit formulae of $g_1$ and $f_1$ in terms of the quantities
appearing in (1.13)
can be expressed explicitly in terms of
$$\{ g_0,f_0,\hat L_1,\hat M_1,\hat M_2, \hat{\LL}_1, \hat{\MM}_1,
  \hat{\MM}_2 \},$$
see equations (4.7) and (4.8).
The Gelfand-Levitan-Marchenko representations imply that $\{ \hat
  L_j,\hat M_j\}^2_{j=1}$, for $t>0$, $k\in {\mathbb{C}}$, satisfy the ODEs
$$ \begin{array}{ll}
\hat L_{1_t} + 4ik^2\hat L_1 = ig_1(t)\hat L_2 + \chi_1(t) \hat M_1 +
\chi_2(t)\hat M_2
+ ig_1(t),\\ \\
\hat L_{2_t} = -i\lambda \overline{g_1}(t) \hat L_1 - \chi_1(t) \hat M_2 +
\lambda \bar
\chi_2(t) \hat M_1, \\ \\
\hat M_{1_t} + 4ik^2\hat M_1 = 2g_0(t) \hat L_2 + ig_1(t) \hat M_2 +
2g_0(t)\\ \\
\hat M_{2_t} = 2\lambda \overline{g_0} (t) \hat L_1 - i\lambda
\overline{g_1} (t) \hat
M_1, \\ \\
\end{array} \eqno (1.16)$$
as well as the initial conditions
$$\hat L_j (0,k) = \hat M_j(0,k) =0, \quad j = 1,2, $$
where the functions $\chi_1(t)$ and $\chi_2(t)$ are defined by
$$\chi_1(t) = \frac{\lambda}{2} (g_0\overline{g_1} - \overline{g_0} g_1),
\quad
\chi_2(t) = \half \frac{dg_0}{dt} - \rho |g_0|^2g_0. \eqno (1.17)$$
$\{ \hat{\LL}_j, \hat{\MM}_j\}^2_{j=1}$ satisfy similar equations with
$g_0,g_1$ replaced by
$f_0,f_1$.

Substituting the expressions for $g_1$ and $f_1$ from (4.7), (4.8)
in equations (1.16) and the analogous
equations for $\{ \hat{\LL}_j, \hat{\MM}_j\}^2_{j=1}$, we obtain a system of
nonlinear
Volterra integral equations for the functions $\{ \hat L_j,\hat M_j,
\hat{\LL}_j,\hat{\MM}_j\}^2_{j=1} $ in terms of $g_0$ and $f_0$.  The rigorous
analysis of
this system remains open.

\paragraph{Organization of the Paper and Notations} \ \

Steps 1-3 are implemented in sections 2-4.

In addition to the notations (1.7) the following notations will also be used
$$ s(k)e^{ikL\hat{\sigma_{3}}}S_L(k)
\equiv s(k)e^{ikL\sigma_{3}}S_L(k)e^{-ikL\sigma_{3}} =
\curl{\overline{\alpha(\bar
k)}}{\beta(k)}{\overline{\lambda
\beta(\bar k)}}{\alpha(k)}. \eqno (1.18)$$

$\displaystyle{\mu^{(*)}}$ denotes a function which is analytic and
bounded for
$\{ k \in {\mathbb{C}}$, arg $k \in L_*\}$, where
$$ L_1: [0,\frac{\pi}{2}], \quad L_2: [\frac{\pi}{2},\pi], \quad L_3: [\pi,
\frac{3\pi}{2}], \quad L_4: [ \frac{3\pi}{2},\pi], \eqno (1.19)$$
$$L_{12}: L_1 \cup L_2, \quad {\mathrm{etc.}}$$

We conclude the introduction with some remarks.

\section{A Riemann-Hilbert Formulation Under the Assumption of Existence}

The NLS equation admits the Lax pair \cite{14} formulation \cite{15}
$$ \mu_x + ik\hat \sigma_3\mu = Q\mu$$
$$ \mu_t + 2ik^2\hat \sigma_3 \mu = \tilde Q\mu, \eqno (2.1)$$
where $\mu(x,t,k)$ is a $2\times 2$ matrix valued function, $\hat \sigma_3$
is defined by
$$ \hat \sigma_3 \cdot = [\sigma_3, \cdot], \quad \sigma_3 = \diag(1,-1),
\eqno (2.2)$$
and the $2\times 2$ matrices $Q$, $\tilde Q$ are defined by
$$ Q(x,t) = \curl{0}{q(x,t)}{\lambda \bar q(x,t),}{0}, \quad \tilde
Q(x,t,k) = 2kQ -iQ_x\sigma_3-i\lambda |q|^2\sigma_3,     \quad \lambda =
\pm 1. \eqno (2.3)$$
The definition of $\hat \sigma_3$ implies that if $A$ is a $2\times 2$
matrix, then
$$ e^{\hat \sigma_3}A = e^{\sigma_3}Ae^{-\sigma_3}. \eqno (2.4)$$

Equations (2.1) can be rewritten as
$$ d\left( e^{i(kx+2k^2t)\hat \sigma_3} \mu(x,t,k)\right) = W(x,t,k), \eqno
(2.5)$$
where the closed 1-form $W$ is defined by
$$ W= e^{i(kx+2k^2t)\hat \sigma_3} (Q\mu dx + \tilde Q\mu dt). \eqno
(2.6)$$

Throughout this section we {\it assume}
 that there exists a sufficiently smooth solution $q(x,t)$, $x\in
[0,L]$, $t \in [0,T]$, of the NLS equation.

A solution of equation  (2.5)
is given by
$$ \mu_*(x,t,k) = I + \int^{(x,t)}_{(x_*,t_*)} e^{-i(kx+2k^2t)\hat
\sigma_3} W(y,\tau,k), \eqno (2.7)$$
where $(x_*,t_*)$  is an arbitrary point in the domain $x \in [0,L]$, $t\in
[0,T]$, and the integral denotes a line integral connecting smoothly the
points indicated.  Following \cite{1} we choose the  point $(x_*,t_*)$ as each of
the corners of the polygonal domain.  Thus we define four different
solutions $\mu_1,...,\mu_4$, corresponding to $(0,T)$, $(0,0)$, $(L,0)$,
$(L,T)$, see figure 1.1

By splitting the line integrals into integrals parallel to the $t$ and the
$x$ axis we find
$$ \mu_2(x,t,k) = I+ \int^x_0 e^{-ik(x-y)\hat \sigma_3} (Q\mu_2)(y,t,k)dy +
e^{-ikx\hat \sigma_3} \int^t_0 e^{-2ik^2 (t-\tau)\hat \sigma_3}(\tilde
Q\mu_2)(0,\tau,k)d\tau, \eqno (2.8)$$

$$ \mu_3(x,t,k) = I- \int^L_x e^{-ik(x-y)\hat \sigma_3}(Q\mu_3)(y,t,k)dy +
e^{-ik(x-L)\hat \sigma_3} \int^t_0 e^{-2ik^2(t-\tau)\hat \sigma_3} (\tilde
Q\mu_3) (L,\tau,k)d\tau, \eqno (2.9)$$
$\mu_1$ and $\mu_4$ satisfy equations similar to those of $\mu_2$ and
$\mu_3$ where $ \int^t_0$ is replaced by $-\int^T_t$.

Note that all the $\mu_{j}$ are entire functions of $k$.

\subsection{Eigenfunctions and Their Relations}

The definitions of $\mu_j$, $j=1,...,4$, and the notations (1.19) imply
$$ \mu_1 = \left( \mu^{(2)}_1, \mu^{(3)}_1\right), \mu_2 = \left(
\mu^{(1)}_2, \mu^{(4)}_2\right), \mu_3 = \left( \mu^{(3)}_3,
\mu^{(2)}_3\right), \mu_4 = \left( \mu^{(4)}_4, \mu^{(1)}_4\right). \eqno
(2.1.1)$$

The functions $\mu_1(0,t,k)$, $\mu_2(0,t.k)$, $\mu_3(x,0,k)$,
$\mu_{3}(L,t.k)$,
 $\mu_4(L,t,k)$ are bounded in
larger domains:

$$\mu_1(0,t,k) = \left( \mu_1^{(24)}(0,t,k), \mu^{(13)}_1(0,t,k)\right),
\quad \mu_2(0,t,k) = \left( \mu^{(13)}_2(0,t,k),
\mu^{(24)}_2(0,t,k)\right),$$
$$\mu_3(x,0,k) = \left( \mu^{(34)}_3(x,0,k), \mu^{(12)}_3(x,0,k)\right),
\quad \mu_3(L,t,k) = \left( \mu^{(13)}_3(L,t,k), \mu^{(24)}_3(L,t,k)\right),
$$
$$ \mu_4(L,t,k) = \left( \mu^{(24)}_4(L,t,k), \mu^{(13)}_4(L,t,k)\right).
\eqno (2.1.2)$$

The matrices $Q$ and $\tilde Q$ are traceless, thus
$$ {\mathrm{det}} \mu_j(x,t,k) =1, \quad j = 1,...,4. \eqno (2.1.3)$$

The definitions of $\mu_j^{(*)}$ imply that in the domains where
these functions are bounded, they satisfy
$$ \mu_j^{(*)} (x,t,k) = I_j^{(*)} + O\left(\frac{1}{k}\right), \quad k
\rightarrow
\infty, \eqno (2.1.4)$$
where vector $I_j^{(*)}$ is either $(0,1)^\dag$ or $(1,0)^\dag$, depending
on what
column of $\mu_{j}$ is made by $\mu_{j}^{(*)}$.

The functions $\mu_j$ are related by the equations
$$ \mu_3(x,t,k) = \mu_2(x,t,k) e^{-i(kx+2k^2t)\hat\sigma_3}s(k), \eqno
(2.1.5)$$
$$ \mu_1(x,t,k) = \mu_2(x,t,k) e^{-i(kx+2k^2t)\hat\sigma_3} S(k), \eqno
(2.1.6)$$
$$ \mu_4(x,t,k) = \mu_3(x,t,k) e^{-i[k(x-L)+ 2ik^2t]\hat\sigma_3}S_L(k).
\eqno
(2.1.7)$$
Evaluating equation (2.1.5) at $x=t=0$ we find $s(k) = \mu_3(0,0,k)$.
Evaluating
equation (2.1.6) at $x=t=0$ we find $S(k) = \mu_1(0,0,k)$; evaluating
equation (2.1.6)
at $x=0$, $t=T$ we find $S(k) = (e^{2ik^2T\hat\sigma_3} \mu_2(0,T,k))^{-1}$.
Evaluating equation (2.1.7) at $x=L$, $t=0$ we find $S_L(k) =
\mu_4(L,0,k)$; evaluating
equation (2.1.6) at $x=L$, $t=T$ we find $S_L(k) =
(e^{2ik^2T\hat\sigma_3} \mu_3(L,T,k))^{-1}$.  Equations (2.1.5) and
(2.1.7) imply
$$\mu_4(x,t,k) = \mu_2(x,t,k) e^{-i(kx+2k^2t)\hat\sigma_3}\Big(
s(k)e^{ikL\hat\sigma_3}
S_L(k)\Big). \eqno (2.1.8)$$
The symmetry properties of $Q$ and $\tilde Q$ imply

$$ (\mu(x,t,k))_{11} = \overline{ (\mu(x,t,\bar k))}_{22}, \quad
(\mu(x,t,k))_{21} = \overline{\lambda(\mu(x,t,\bar k))}_{12}. \eqno
(2.1.9)$$

The definitions of $\mu_3(0,0,k)$, $\mu_2(0,T,k)$, $\mu_3(L,0,k)$
imply
$$ s(k) = I - \int^L_0 e^{iky\hat \sigma_3}(Q\mu_3) (y,0,k)dy, \eqno
(2.1.10)$$
$$ S^{-1}(k) = I+ \int^T_0 e^{2ik^2\tau \hat \sigma_3}(\tilde
Q\mu_2)(0,\tau,k)d\tau, \eqno (2.1.11)$$
$$ S^{-1}_L(k) = I+ \int^T_0 e^{2ik^2\tau\hat \sigma_3}(\tilde
Q\mu_3)(L,\tau,k)d\tau. \eqno (2.1.12)$$

The symmetry conditions (2.19) justify the notations (1.7).

Equations (2.1.2), the determinant condition (2.1.3), and the large $k$
behaviour of $\mu_j$ imply the following properties:

{\it{\underline{$a(k),b(k)$}}}

\begin{itemize}

\item $a(k)$, $b(k)$ are entire functions.

\item $a(k)\overline{a(\bar k)} - \lambda b(k) \overline{b(\bar k)} =1$, $k
\in {\mathbb{C}}$.

\item $$\displaystyle{ a(k) = 1 + O\left( \frac{1+e^{2ikL}}{k}\right), \quad
b(k) = O\left( \frac{1+e^{2ikL}}{k}\right), \quad k \rightarrow \infty;}$$
in particular,
$$a(k),\quad b(k),\quad \overline{a(\bar k)}e^{2ikL}, \quad
\overline{b(\bar k)} e^{2ikL}
\quad \mbox{are bounded for}\quad \arg k \in \left[ 0, \pi \right].
\eqno(2.1.13)$$

\end{itemize}

{\it{\underline{$A(k), B(k)$}}}

\begin{itemize}
\item $A(k)$, $B(k)$ are  entire functions.

\item $A(k) \overline{A(\bar k)} - \lambda B(k) \overline{B(\bar k)} =1$,
$k \in {\mathbb{C}}$.

\item $$\displaystyle{ A(k) = 1 + O\left( \frac{1 + e^{4ik^2T}}{k} \right),
\quad B(k) = O\left( \frac{1+e^{4ik^2T}}{k}\right), \quad k \rightarrow
\infty;} \eqno(2.1.14)$$
in particular,
$$A(k),\quad B(k)
\quad \mbox{are bounded for}\quad \arg k \in \left[ 0,
\frac{\pi}{2}\right] \cup
[\pi, \frac{3\pi}{2}] .
$$
\end{itemize}

\vskip .15in
\noindent
{\it{\underline{ $\A(k)$, $\B(k)$ }}}

Same as $A(k)$, $B(k)$.

\subsection{The Global Relation}

{\bf Proposition 2.1.}
{\it Let the spectral functions $a(k)$, $b(k)$, $A(k)$, $B(k)$, $\A(k)$,
$\B(k)$, be defined in equations (1.7), where $s(k)$, $S(k)$,
$S_L(k)$ are defined by equations (1.6), and $\mu_2,\mu_3$ are defined by
equations
(2.8), (2.9) in terms of the smooth function $q(x,t)$.
These spectral functions are not independent but they satisfy
the global relation (1.4) where $c^+(k)$ denotes the (12) element of
$-\int^L_0 [\exp(iky\hat \sigma_3)] (Q\mu_4)(y,T,k)dy$, and $\mu_{4}$
is defined by an equation similar to $\mu_{3}$ with $\int_{0}^{t}$
replaced by $-\int_{t}^{T}$.
\vskip .2in
}
\noindent {\bf Proof. }

Evaluating equation (2.1.8) at $x=0$, $t=T$, and writing $\mu_2(0,T,k)$ in
terms of
$S(k)$ we find
$$\mu_4(0,T,k) = e^{-2ik^2T\hat\sigma_3}\left(
S^{-1}(k)s(k)e^{ikL\hat\sigma_3}S_L(k)
\right).$$
Multiplying this equation by $\exp[2ik^2T\hat\sigma_3]$ and using the
definition of
$\mu_4(x,T,k)$ we find
$$ -I + S^{-1}s \left( e^{ikL \hat\sigma_3}S_L\right) +
e^{2ik^2T\hat\sigma_3} \int^L_0 e^{iky\hat\sigma_3} (Q\mu_4)(y,T,k)dy
= 0. $$
The (12) element of this equation is equation (1.4).

\subsection{The Jump Conditions}

Let $M(x,t,k)$ be defined by
$$ M_+ = \left( \frac{\mu^{(1)}_2}{\alpha(k)}, \mu^{(1)}_4\right), \arg k \in
\left[ 0, \frac{\pi}{2}\right]; M_- = \left( \frac{\mu^{(2)}_1}{d(k)},
\mu^{(2)}_3\right), \arg k\in \left[ \frac{\pi}{2}, \pi\right]; $$
$$ M_+= \left( \mu^{(3)}_3, \frac{\mu^{(3)}_1}{\overline{d(\bar
k)}}\right), \arg k \in \left[ \pi, \frac{3\pi}{2}\right]; M_- = \left(
\mu^{(4)}_4, \frac{\mu_2^{(4)}}{\overline{\alpha(\bar k)}}\right), \arg
k\in \left[ \frac{3\pi}{2}, 2\pi\right]; \eqno (2.3.1)$$
where the scalars $d(k)$ and $\alpha(k)$ are defined below, see (2.3.7),
(2.3.8).

These definition imply
$$ \det M(x,t,k) =1, \eqno (2.3.2)$$
and
$$M(x,t,k) = I+O\left( \frac{1}{k}\right), \quad k\rightarrow \infty. \eqno
(2.3.3)$$

\noindent{\bf Proposition 2.2}.  {\it Let $M(x,t,k)$ be defined by
equations (2.3.1), where $\mu_2(x,t,k)$, $\mu_3(x,t,k)$ are defined by
equation (2.8), (2.9), $\mu_1(x,t,k)$, $\mu_4(x,t,k)$ are defined by
similar equations with $\int_{0}^{t}$ replaced by $-\int_{t}^{T}$ and
$q(x,t)$ is a smooth function.  Then $M$ satisfies the ``jump'' condition
$$ M_-(x,t,k) = M_+(x,t,k)J(x,t,k), \quad k \in {\mathbb{R}} \cup i
{\mathbb{R}},
\eqno (2.3.4)$$
where the $2\times 2$ matrix $J$ is defined by
$$J = \left[ \begin{array}{ll}
J_2, & \arg k =0 \\ \\
J_1, & \arg k = \frac{\pi}{2} \\ \\
J_{4} \equiv J_3J^{-1}_2J_1, & \arg k = \pi \\ \\
J_3 & \arg k = \frac{3\pi}{2}, \end{array} \right. \eqno (2.3.5)$$
and
$$ J_1 = \left( \begin{array}{lc}
\frac{\delta(k)}{d(k)} & -\B (k)e^{2ikL}e^{-2i\theta} \\ \\
\frac{\lambda \overline{B(\bar k)}}{d(k)\alpha(k)}e^{2i\theta}  &
\frac{a(k)}{\alpha(k)} \end{array} \right),
J_3 = \curl{
\frac{\overline{\delta(\bar k)}}{\overline{d(\bar
k)}}}{\frac{-B(k)}{\overline{d(\bar k)} \overline{\alpha(\bar
k)}}e^{-2i\theta}}{\lambda \overline{\B(\bar k)}e^{-2ikL}e^{2i\theta}}{
\frac{\overline{a(\bar k)}}{\overline{\alpha(\bar k)}}}, $$
$$ J_2 = \left( \begin{array}{cc}
1 & -\frac{\beta(k)}{\overline{\alpha( k)}}e^{-2i\theta} \\ \\
\lambda \frac{\overline{\beta( k)}}{\alpha(k)}e^{2i\theta} &
\frac{1}{|\alpha(k)|^2} \end{array}\right), \quad \theta (x,t,k) =  kx +
2 k^2t. \eqno (2.3.6)$$
$$ \alpha(k) = a(k)\A(k) + \lambda \overline{b(\bar k)}e^{2ikL}\B(k),
\qquad \beta(k) = b(k)\A(k) + \overline{a(\bar k)}e^{2ikL}\B(k), \eqno
(2.3.7)$$
$$d(k) = a(k)\overline{A(\bar k)}-\lambda b(k) \overline{B(\bar k)}, \quad
\delta(k) = \alpha(k)\overline{A(\bar k)} - \lambda
\beta(k)\overline{B(\bar k)}. \eqno (2.3.8)$$
}

\noindent {\bf Proof.
}

Writing equations (2.1.8), (2.1.5), (2.1.6) in vector form we find
$$ \mu^{(4)}_4 = \bar \alpha \mu^{(1)}_2 + \lambda \bar \beta e\mu^{(4)}_2,
\quad \mu^{(1)}_4 = \beta \bar e\mu^{(1)}_2 + \alpha\mu^{(4)}_2; \eqno
(2.3.9)$$
$$ \mu^{(3)}_3 = \bar a \mu^{(1)}_2 + \lambda \bar b e\mu^{(4)}_2, \quad
\mu^{(2)}_3 = b\bar e\mu^{(1)}_2 + a\mu^{(4)}_2; \eqno (2.3.10)$$
$$ \mu^{(2)}_1 = \bar A\mu^{(1)}_2 + \lambda \bar B e\mu^{(4)}_2, \quad
\mu^{(3)}_1 = \bar B\bar e \mu^{(1)}_2 + A\mu^{(4)}_2, \eqno (2.3.11)$$
where $e = \exp (2i\theta)$.
Recall that $\alpha$ and $\beta$ are the (22) and (12) elements of
$se^{ikL\hat{\sigma}_{3}}S_L$ (see (1.18)), thus
$$ \alpha(k)\overline{\alpha(\bar k)} - \lambda \beta(k)
\overline{\beta(\bar k)} =1. \eqno (2.3.12)$$

Rearranging equations (2.3.9) and using equation (2.3.12) we find the jump
condition across $\arg k =0$.

In order to derive the jump condition across $\arg k = \frac{\pi}{2}$, we
first eliminate $\mu^{(4)}_2$ from equations (2.3.9b) and (2.3.11a):
$$  \mu_1^{(2)} = \frac{\delta \mu^{(1)}_2}{\alpha} + \frac{\lambda\bar
Be}{\alpha} \mu^{(1)}_4. \eqno (2.3.13a)$$
We then eliminate $\mu^{(4)}_2$ from equations (2.3.9b) and (2.3.10b):
$$ \mu_3^{(2)} = (b\alpha - a\beta) \frac{\bar e\mu^{(1)}_2}{\alpha} +
\frac{a\mu^{(1)}_4}{\alpha}. \eqno (2.3.13b)$$
Using the identity
$$a \beta - b\alpha = e^{2ikL}\B, \eqno (2.3.14)$$
and dividing equation (2.3.13a) by $d$, equations (2.3.13) define the jump
across $\arg k = \frac{\pi}{2}$.

The jump across $\arg k = \frac{3\pi}{2}$ follows from symmetry
considerations and the jump across $\arg k = \pi$ follows from the fact
that the product of the jump matrices must equal the identity.

We note that the jump matrices have unit determinant; in particular
regarding $J_1$ we note that
$$ d\alpha - \lambda \bar B\B e^{2ikL} = a\delta; \eqno (2.3.15)$$
indeed, the lhs of this equation equals
$$ \alpha(a\bar A - \lambda b\bar B) -  \lambda \bar B\B e^{2ikL} = \alpha
a\bar A-\lambda \bar B(\B e^{2ikL} + \alpha b) = $$
$$  = \alpha a \bar A - \lambda \bar B a\beta = a\delta, $$
where we have used the identity (2.3.14).

\subsection{The Residue Relations}

{\bf Proposition 2.3}
{\it Let $\alpha(k)$ and $d(k)$ be defined by equations (2.3.7) and (2.3.8)
in terms of the spectral functions considered in proposition 2.1.  Assume
that:

\begin{itemize}
\item $\alpha(k)$ has simple zeros, $\{ \nu_j\}$, $\arg \nu_j \in
(0,\frac{\pi}{2})$, and has {\bf no} zeros for $\arg k =0$  and $\arg k =
\frac{\pi}{2}$.
\item $d(k)$ has simple zeros, $\{ \lambda_j \}$, $\arg \lambda_j \in
\left( \frac{\pi}{2},\pi\right)$, and has {\bf no} zeros for $\arg k =
\frac{\pi}{2}$ and $\arg k=\pi$.
\item None of the zeros of $d(k)$ for $\arg k \in
(\frac{\pi}{2},\pi )$, coincides with any of the zeros of $a(k)$.
\item None of the zeros of $\alpha(k)$ for $\arg k \in
(0,\frac{\pi}{2})$, coincides with any of the zeros of $a(k)$.

\end{itemize}

Let $[M]_1$ and $[M]_2$ denote the first and the second column of the
matrix $M$.  Then
$$
 Res_{k=\nu_j}
 [M(x,t,k)]_1 = c^{(1)}_j e^{4i\nu^2_jt + 2i\nu_jx}
[M(x,t,\nu_j)]_2, \eqno (2.4.1)$$
$$
 Res_{k=\bar \nu_j}
[M(x,t,k)]_2 = \lambda \overline{c^{(1)}_j}
e^{-4i\bar \nu^2_jt-2i\bar \nu_jx} [M(x,t,\bar \nu_j)_1, \eqno (2.4.2)$$
$$
 Res_{k=\lambda_j}
 [M(x,t,k)]_1 = c^{(2)}_j e^{4i\lambda^2_jt
+2i\lambda_jx}[M(x,t,\lambda_j)]_2, \eqno (2.4.3)$$
$$
 Res_{k=\bar \lambda_j}
 [M(x,t,k)]_2 = \lambda
\overline{c^{(2)}_j}e^{-4i\bar{\lambda}^2_jt-2i\bar{\lambda}_jx}[M(x,t,\bar\lambda_j)]_1,
\eqno (2.4.4) $$
where
$$ c^{(1)}_j = \frac{a(\nu_j)}{e^{2i\nu_jL}\B(\nu_j)\dot \alpha(\nu_j) },
\quad c^{(2)}_j = \frac{\lambda \overline{B(\bar
\lambda_j)}}{a(\lambda_j)\dot d(\lambda_j)}. \eqno (2.4.5)$$
}

\noindent {\bf Proof.
}

Let us first notice that the assumptions of the proposition, the
definitions of the quantities $\alpha(k)$ and $d(k)$ (see (2.3.7)
and (2.3.8)) and the identities,
$$
A(k)\overline{A(\bar k)} - \lambda B(k)\overline{B(\bar k)} = 1,
\quad
\A(k)\overline{\A(\bar k)} - \lambda \B(k)\overline{\B(\bar k)} = 1,
$$
imply that
$$
B(\bar{\lambda_{j}}) \neq 0\quad \mbox{and}\quad  \B(\nu_{j}) \neq 0,
$$
so that
the rhs of equations (2.4.5) are well defined. We proceed now with the
proof of the proposition.

The matrix $J_1(x,t,k)$ defined in equations (2.3.6) can be factorized as
follows:

$$J_1(x,t,k) =
\curl{\frac{\alpha(k)}{a(k)}}{-\B(k)e^{2ikL}e^{-2i\theta}}{0}{
\frac{a(k)}{\alpha(k)}} \curl{1}{0}{ \frac{\lambda \overline{B(\bar
k)}e^{2i\theta}}{a(k)d(k)}}{1}. \eqno (2.4.6)$$

Indeed, the entries (12), (21), (22) are equal identically.  The entries
(11) are equal iff
$$ \delta a = \alpha d - \lambda \B \overline{B(\bar k)} e^{2ikL}.$$
Replacing in this equation $\delta$ and $d$ by their definitions (see
equations (2.3.8)) we find the identity (2.3.14).

Using the factorization (2.4.6), the jump condition $M_- = M_+J_1$ becomes
$$ \left( \frac{\mu^{(2)}_1}{d}, \mu^{(2)}_3\right) \curl{1}{0}{
\frac{-\lambda\bar B e^{2i\theta}}{ad}}{1} = \left(
\frac{\mu^{(1)}_2}{\alpha}, \mu^{(1)}_4\right) \curl{\frac{\alpha}{a}}{- \B
e^{2ikL}e^{-2i\theta}}{0}{\frac{a}{\alpha}}. \eqno (2.4.7)$$
Evaluating the second column of this equation at $k = \nu_j$
(we remind that all the functions involved are entire) we find
$$ 0 = -\B(\nu_j)e^{2i\nu_jL}e^{-2i\theta(\nu_j)} \mu^{(1)}_2(\nu_j) +
a(\nu_j) \mu_4^{(1)}(\nu_j), \eqno (2.4.8)$$
where for convenience of notation we have suppressed the $x,t$ dependence
of $\mu^{(1)}_2$, $\mu^{(1)}_4$, $\theta$.  Hence,

$$
 Res_{k=\nu_j}
 [M(x,t,k)]_1 = \frac{\mu^{(1)}_2(x,t,\nu_j)}{ \dot
\alpha(\nu_j)} = \frac{a(\nu_j)e^{2i\theta(x,t,\nu_j)}
\mu^{(1)}_4(x,t,\nu_j)}{e^{2i\nu_jL}\B(\nu_j)\dot \alpha(\nu_j)}, $$
which, using $\mu^{(1)}_4(x,t,\nu_j) = [M(x,t,\nu_j)]_2$, becomes equation
(2.4.1).

Similarly, evaluating the first column of equation (2.4.7) at $k =
\lambda_j$ we find

$$ 0 = \mu^{(2)}_1(\lambda_j) - \frac{\lambda \overline{B(\bar \lambda_j)}
e^{2i\theta(\lambda_j)}}{a(\lambda_j)} \mu^{(2)}_3 (\lambda_j).$$
Hence

$$
 Res_{k=\lambda_j}
 [M(x,t,k)]_1 =
\frac{\mu^{(2)}_1(x,t,\lambda_j)}{\dot d(\lambda_j)} = \frac{\lambda
\overline{B(\bar
\lambda_j)}e^{2i\theta(x,t,\lambda_j)\mu^{(2)}_3(x,t,\lambda_j)}}{a
(\lambda_j)\dot d(\lambda_j)},$$
which yields (2.4.3).

Equations (2.4.2) and (2.4.4) follow from equations (2.4.1) and (2.4.3)
using symmetry considerations.

\section{Existence Under the Assumption that the Global Relation is Valid}

\subsection{The Spectral Functions}

The analysis of \S 2 motivates the following definitions and results for
the spectral functions.  The relevant rigorous analysis can be found in \cite{6}.

\vskip .2in
\noindent {\bf Definition 3.1}  {\it (The spectral functions $a(k)$, $b(k)$
)  Given the smooth function $q_0(x)$, define the vector $\phi(x,k) =
(\phi_1,\phi_2)^\dag$ as the unique solution of
$$ \phi_{1_x} + 2ik\phi_1 = q_0(x)\phi_2, \phi_{2_x} = \lambda
\bar q_0(x)\phi_1, 0<x<L, k\in {\mathbb{C}}, \phi(L,k) = (0,1)^\dag.
\eqno (3.1)$$
Given $\phi(x,k)$ define the functions $a(k)$ and $b(k)$ by
$$ a(k) = \phi_2(0,k), \quad b(k) = \phi_1(0,k), \quad k \in
{\mathbb{C}}. \eqno (3.2)$$
}

\noindent {\bf Properties of $a(k)$, $b(k)$
}

\begin{itemize}
\item $a(k)$, $b(k)$ are entire functions.
\item $a(k)\overline{a(\bar k)} - \lambda b(k) \overline{b(\bar k)} =1$, $k
\in {\mathbb{C}}$.
\item $$\displaystyle{ a(k) = 1 + O\left( \frac{1+e^{2ikL}}{k}\right), \quad
b(k) = O\left( \frac{1+e^{2ikL}}{k}\right), \quad k \rightarrow \infty;}$$
in particular,
$$a(k),\quad b(k),\quad \overline{a(\bar k)}e^{2ikL}, \quad
\overline{b(\bar k)} e^{2ikL}
\quad \mbox{are bounded for}\quad \arg k \in \left[ 0, \pi \right].
\eqno(2.1.13)$$
\end{itemize}
We shall also assume that $a(k)$ has at most simple zeros, $\{ k_j\}$, for
Im $k_j >0$
and has no zeros for Im $k=0$.

\noindent {\bf Remark 3.1}  The definition 3.1 gives rise to the map,
$${\mathbb{S}}:  \{ q_0(x)\} \rightarrow \{ a(k),b(k) \}. \eqno (3.4a)$$
The inverse of this map,
$$ {\mathbb{Q}}: \{ a(k), b(k) \} \rightarrow \{ q_0(x)\}, \eqno (3.4b)$$
can be defined as follows:
$$ q_0(x) = 2i\lim_{k\rightarrow\infty} (kM^{(x)}(x,k))_{12}, \eqno (3.5)$$
where $M^{(x)}(x,k)$ is the unique solution of the following RH problem:
\begin{itemize}
\item
$$ M^{(x)}(x,k) = \left[ \begin{array}{ll}
M^{(x)}_-(x,k), & {\mathrm{Im}} k \leq 0 \\ \\
M^{(x)}_+(x,k), & {\mathrm{Im}} k \geq 0, \end{array} \right. $$
is a sectionally meromorphic function with unit determinant.

\item
$$ M^{(x)}_-(x,k) = M^{(x)}_+(x,k) J^{(x)}(x,k), \quad k \in
{\mathbb{R}},$$
where

$$ J^{(x)}(x,k) = \curl{1}{- \frac{b(k)}{\bar a(k)}e^{-2ikx}}{
\frac{\lambda\bar b(k)e^{2ikx}}{a(k)}}{\frac{1}{|a|^2}}.$$
\item
$$M^{(x)}(x,k) = I + O\left(\frac{1}{k}\right), \quad k
\rightarrow \infty.$$

\item
The first column of $M^{(x)}_+$ can have simple poles at $k=k_j$,
and the second column of $M^{(x)}_-$ can have simple poles at $k = \bar
k_j$, where $\{ k_j\}$ are the simple zeros of $a(k)$, Im $k_j >0$.  The
associated residues are given by
$$
 Res_{k=k_j}
[M^{(x)}(x,k)]_1 = \frac{e^{2ik_jx}}{\dot
a(k_j)b(k_j)} [M^{(x)}(x,k_j)]_2, $$
$$
 Res_{k=\bar k_j}
 [M^{(x)}(x,k)]_2 = \frac{\lambda
e^{-2i{\bar k}_jx}}{\overline{\dot a(k_j)} \overline{b(k_j)}}
[M^{(x)}(x,\bar k_j)]_1. \eqno (3.6)$$

It can be shown (see for example \cite{6}) that

$$ {\mathbb{S}}^{-1} = {\mathbb{Q}}. \eqno (3.4c)$$
\end{itemize}

\noindent{\bf Definition 3.2} {\it (The spectral functions $A(k)$, $B(k)$).
Let
$$ Q^{(0)}(t,k) = 2k \curl{0}{g_0(t)}{\lambda\bar g_0(t)}{0} - i
\curl{0}{g_1(t)}{\lambda \bar g_1(t)}{0} \sigma_3 -i\lambda
| g_0(t)|^2\sigma_3, \lambda = \pm 1. \eqno (3.7)$$
Given the smooth functions $g_0 (t)$, $g_1(t)$, define the vector
$\Phi(t,k) = (\Phi_1,\Phi_2)^\dag$ as the unique solution of
$$ \Phi_{1_t} + 4ik^2\Phi_1 = Q^{(0)}_{11} \Phi_1 + Q^{(0)}_{12}\Phi_2,$$
$$ \Phi_{2_t} = Q^{(0)}_{21} \Phi_1 + Q^{(0)}_{22} \Phi_2, \quad 0<t<T,
\quad k\in {\mathbb{C}}, $$
$$ \Phi(0,k) = (0,1)^\dag. \eqno (3.8)$$
Given $\Phi(t,k)$ define the functions $A(k)$ and $B(k)$ by
$$ A(k) = \overline{\Phi_2(T,\bar k)}, \quad B(k) = -\Phi_1(T,k)
e^{4ik^2T}.
\eqno (3.9)$$
}

\noindent {\bf Properties of $A(k)$, $B(k)$
}
\begin{itemize}
\item $A(k)$, $B(k)$ are  entire functions.

\item $A(k) \overline{A(\bar k)} - \lambda B(k) \overline{B(\bar k)} =1$,
$k \in {\mathbb{C}}$.

\item $$\displaystyle{ A(k) = 1 + O\left( \frac{1 + e^{4ik^2T}}{k} \right),
\quad B(k) = O\left( \frac{1+e^{4ik^2T}}{k}\right), \quad k \rightarrow
\infty;} \eqno(2.1.14)$$
in particular,
$$A(k),\quad B(k)
\quad \mbox{are bounded for}\quad \arg k \in \left[ 0,
\frac{\pi}{2}\right] \cup
[\pi, \frac{3\pi}{2}] .
$$
\end{itemize}
We shall also assume that $A(k)$  has at most simple zeros, $\{ K_j\}$, for
arg $K_j \in \left(0, \frac{\pi}{2}\right)\cup \left(\pi,
\frac{3\pi}{2}\right)$
and has no zeros for arg $k=0$, $\frac{\pi}{2}$, $\pi$,  $\frac{3\pi}{2}$ .

\noindent {\bf Remark 3.2} The definition 3.2 gives rise to the map

$$ {\mathbb{S}}^{(0)}  : \{ g_0(t), g_1(t)\} \rightarrow \{ A(k), B(k) \}.
\eqno
(3.11a)$$
The inverse of this map
$$ \mathbb{Q}^{(0)}: \{ A(k), B(k) \} \rightarrow \{ g_0(t), g_1(t)\}, \eqno
(3.11b)$$
can be defined as follows:
$$ g_0(t) = 2i\lim_{k\rightarrow \infty} (kM^{(0)}(t,k))_{12}, $$
$$ g_1(t) = \lim_{k\rightarrow \infty} \left\{ 4(k^2M^{(0)}(t,k))_{12} +
2ig_0(t)k(M^{(0)}(t,k) -I)_{22}  \right\},$$
where $M^{(0)}(t,k)$ is the unique solution of the following RH problem:

\begin{itemize}
\item
$$ M^{(0)} (t,k) = \left[ \begin{array}{ll}
M^{(0)}_+(t,k), & \arg k \in [0, \frac{\pi}{2}] \cup [\pi, \frac{3\pi}{2}]
\\ \\
M^{(0)}_-(t,k), & \arg k \in [\frac{\pi}{2}, \pi] \cup [\frac{3\pi}{2},
2\pi], \end{array} \right. $$
is a sectionally meromorphic function with unit determinant.

\item
$$M^{(0)}_-(t,k) = M^{(0)}_+(t,k) J^{(0)}(t,k), \quad k \in
{\mathbb{R}} \cup i{\mathbb{R}},$$
where
$$ J^{(0)}(t,k) = \curl{1}{- \frac{B(k)}{\overline{A(\bar k)}}
e^{-4ik^2t}}{ \frac{\lambda \overline{B(\bar k)}e^{4ik^2t}}{A(k)}}{
\frac{1}{A(k)\overline{A(\bar k)}}}.$$

\item
$$M^{(0)}(t,k)= I+O(\frac{1}{k}), \quad k \rightarrow
\infty.$$

\item  The first column of $M^{(0)}_+(t,k)$ can have simple poles at $k
= K_j$, and the second column of $M^{(0)}_-(t,k)$ can have
simple poles at $k = \bar K_j$, where $K_j  $ are the simple zeros of
$A(k)$, arg $k \in (0,\frac{\pi}{2}) \cup (\pi, \frac{3\pi}{2})$.  The
associated residues are given by
\end{itemize}

$$
 Res_{k=K_j}
 [M^{(0)}(t,k)]_1 = \frac{\exp[4iK^2_jt]}{\dot
A(K_j)B(K_j)} [M^{(0)}(t,K_j)]_2, $$

$$
 Res_{k=\bar K_j}
[M^{(0)}(t,k)]_2 = \frac{\lambda \exp[-4i\bar
K^2_jt]}{\overline{\dot A( K_j)} \overline{B( K_j)}}
[M^{(0)}(t,\bar K_j)]_1.$$

It can be shown, see again \cite{6}, that
$$ \left( {\mathbb{S}}^{(0)} \right)^{-1} = \mathbb{Q}^{(0)}. \eqno (3.11c)$$

\noindent {\bf Definition 3.3} {\it  (The spectral functions $\A(k), \B(k)$).
Let $Q^{(L)}(t,k)$ be defined by an equation similar to (3.7) with
$g_0(t)$, $g_1(t)$ replaced by $f_0(t)$, $f_1(t)$.  Given the smooth
functions $f_0(t)$, $f_1(t)$ define the vector $\varphi (t,k)$ by equations
similar to (3.8) with $Q^{(0)}(t,k)$ replaced by $Q^{(L)}(t,k)$.  Given
$\varphi (t,k)$ define $\A(k)$ and $\B(k)$ by
$$ \A(k) = \overline{\varphi_2(T,\bar k)}, \B(k) = -\varphi_1(T,k)e^{4ik^2T}.
\eqno
(3.14)$$
}

\noindent {\bf Properties of $\A(k)$, $\B(k)$
}

Identical to those of $A(k)$, $B(k)$.
We will denote the zeros of $\A(k)$ by ${\K}_j$.
\vskip .2in

\noindent {\bf Remark 3.3}  The maps
\vskip .15in

\ \ $\qquad \ \ \qquad   \mathbb{S}^{(L)}$: $\{  f_0(t), f_1(t)  \}
\rightarrow \{ {\A}(k),{\B}(k) \}$, \hfill (3.15a)
\vskip .1in

\noindent and

 \ \ $\qquad \ \ \qquad  \QQ^{(L)}$: $ \{ {\A}(k),{\B}(k)\} \rightarrow \{
 f_0(t),f_1(t)  \}$,  \hfill (3.15b)
\vskip .1in

\noindent are defined exactly as in Remark 3.2, where we use the notations
$$M^{(L)}(t,k), J^{(L)}(t,k), {\mathcal{K}}_j, {\mathrm{instead \ \ of}} \
\ M^{(0)}(t,k), J^{(0)}(t,k), K_j. \eqno (3.16)$$
In analogy with equation (3.11c) we find
$$ \left({\mathbb{S}}^{(L)}\right)^{-1} = \QQ^{(L)}. \eqno (3.15c)$$

{\bf Definition 3.4}  {\it (An admissible set).  Given the smooth function
$q_0(x)$ define $a(k)$, $b(k)$ according to definition 3.1.  Suppose that
there exist smooth functions $g_0(t)$, $g_1(t)$, $f_0(t)$, $f_1(t)$, such
that:

\begin{itemize}
\item The associated $A(k)$, $B(k)$, $\A(k)$, $\B(k)$, defined according to
definitions 3.2 and 3.3, satisfy the relation
$$ (a\A + \lambda \bar be^{2ikL}\B)B - (b\A + \bar a e^{2ikL}\B)A =
e^{4ik^2t}c^+(k), k\in {\mathbb{C}}, \eqno (3.16)$$
where $c^+(k)$ is an entire function, which is bounded for Im $k\geq 0$ and
$c^+(k) = O\left(\frac{1+ e^{2ikL}}{k}\right)$, $k\rightarrow \infty$.

\item
$$ g_0(0) = q_0(0), g_1(0) = q'_0(0), f_0(0) = q_0(L), f_1(0) = q'_0(L).
\eqno (3.17)$$
Then we call the functions $g_0(t), g_1(t), f_0(t),f_1(t)$, an admissible
set of functions with respect to $q_0(x)$.
\end{itemize}
}

\subsection{The Riemann-Hilbert Problem}

{\bf Theorem 3.1}  {\it Let $q_0(x)$ be a smooth function.  Suppose that
the set of functions $g_0(t)$, $g_1(t)$, $f_0(t)$, $f_1(t)$, are admissible
with respect to $q_0(x)$, see definition 3.4.  Define the spectral
functions $a(k)$, $b(k)$,  $A(k) $, $B(k)$, $\A(k)$, $\B(k)$, in terms of
$q_0(x)$, $g_0(t)$, $g_1(t)$, $f_0(t)$, $f_1(t)$, according to definitions
3.1, 3.2, 3.3.  Assume that:

\begin{itemize}
\item $a(k)$ has at most simple zeros, $\{ k_j\}$, for Im $k_j >0$ and has
{\bf no} zeros for Im $k=0$.

\item  $A(k)$  has at most simple zeros, $\{ K_j\}$, for
arg $K_j \in \left(0, \frac{\pi}{2}\right)\cup \left(\pi,
\frac{3\pi}{2}\right)$
and has {\bf no} zeros for arg $k=0$, $\frac{\pi}{2}$, $\pi$,
$\frac{3\pi}{2}$ .

\item  $\A(k)$  has at most simple zeros, $\{ \K_j\}$, for
arg $\K_j \in \left(0, \frac{\pi}{2}\right)\cup \left(\pi,
\frac{3\pi}{2}\right)$
and has {\bf no} zeros for arg $k=0$, $\frac{\pi}{2}$, $\pi$,
$\frac{3\pi}{2}$ .

\item The function
$$ d(k) = a(k) \overline{A(\bar k)} - \lambda b(k) \overline{B(\bar k)},
\eqno (3.18)$$
has at most simple zeros, $\{ \lambda_j\}$, for $\arg \lambda_j
\in (\frac{\pi}{2}, \pi)$  and has {\bf no} zeros for $\arg k =
\frac{\pi}{2}$ and $\arg k =\pi$.

\item The function
$$ \alpha(k) = a(k){\A}(k) + \lambda \overline{b(\bar k)}e^{2ikL}\B(k),
\eqno (3.19)$$
has at most simple zeros, $\{ \nu_j\}$, for $\arg \nu_j
\in (0,\frac{\pi}{2})$  and has {\bf no} zeros for $\arg k = 0$
$\arg k= \frac{\pi}{2}$.

\item None of the zeros of $a(k)$ for $\arg k \in (\frac{\pi}{2},\pi)$,
coincides with a zero of $d(k)$.

\item None of the zeros of $a(k)$ for $\arg k \in (0,\frac{\pi}{2}) $,
coincides with a
zero of $\alpha (k)$.

\item None of the zeros of $\alpha(k)$ for $\arg k \in (0,\frac{\pi}{2})
$, coincides with a
zero of $A (k)$ or a zero of $\A(k)$.

\item None of the zeros of $d(k)$ for $\arg k \in (\frac{\pi}{2}, \pi) $,
coincides with a
zero of $\overline{A (\bar k)}$ or a zero of $\overline{\A(\bar k)}$.
\end{itemize}

Define $M(x,t,k) $ as the solution of the following $2 \times 2$ matrix RH
problem:

\begin{itemize}
\item $M$ is sectionally meromorphic in ${\mathbb{C}}\slash \{
{\mathbb{R}} \cup
i{\mathbb{R}} \}$, and has unit determinant.

\item $$ M_-(x,t,k) = M_+(x,t,k)J(x,t,k), \quad k \in {\mathbb{R}} \cup
i{\mathbb{R}}, \eqno (3.20)$$
where $M$ is $M_-$ for $\arg k \in [\frac{\pi}{2},\pi] \cup
[\frac{3\pi}{2}, 2\pi]$, $M$ is $M_+$ for $\arg k \in [0, \frac{\pi}{2}]
\cup [\pi, \frac{3\pi}{2}]$, and $J$ is defined in terms of $a,b,A,B$,
$\A,\B$, by equations (2.3.5), (2.3.6).

\item $$M(x,t,k) = I + O\left( \frac{1}{k}\right), \quad k \rightarrow
\infty. \eqno (3.21)$$

\item Residue conditions (2.4.1) - (2.4.5).

\end{itemize}

Then $M(x,t,k)$ exists and is unique.

Define $q(x,t)$ in terms of $M(x,t,k)$ by
$$q(x,t) = 2i\lim_{k\rightarrow\infty} k(M(x,t,k))_{12}. \eqno (3.22)$$
Then $q(x,t)$ solves the NLS equation (1.1) with
$$ q(x,0) = q_0(x), q(0,t) = g_0(t), q_x(0,t) = g_1(t), q(L,t) = f_0(t),
q_x(L,t) = f_1(t). \eqno (3.23)$$
}

{\bf Proof. }

If $\alpha(k)$ and $d(k)$ have no zeros for arg
$k \in \left(0, \frac{\pi}{2}\right)$ and  for arg
$k \in \left(\frac{\pi}{2}, \pi\right)$ respectively,
then the function $M(x,t,k)$ satisfies a
non-singular RH problem.  Using the fact that the jump matrix $J$
satisfies appropriate
symmetry conditions it is possible to show that this problem has a unique
global solution \cite{16}.  The case that $\alpha(k)$ and $d(k)$ have a finite
number of zeros can be mapped to the case of no zeros supplemented by an
algebraic system of equations which is always uniquely solvable \cite{16}.

{\bf Proof that $q(x,t)$ satisfies the NLS}

Using arguments of the dressing method \cite{2}, it can be verified directly that
if $M(x,t,k)$ is defined as the unique solution of the above RH problem,
and if $q(x,t)$ is defined in terms of $M$ by equation (3.22), then $q$ and
$M$ satisfy both parts of the Lax pair, hence $q$ solves the NLS equation.

{\bf Proof that $q(x,0) = q_0(x)$.}

Let the $2\times 2$ matrices $\hat J_1(x,k)$, $\hat J_3(x,k)$,
$J^{(\infty)}_1(x,k)$, $J^{(\infty)}_2(x,k)$, $J^{(\infty)}_3(x,k)$, be
defined by
$$ \hat J_1 = \curl{ \frac{\alpha(k)}{a(k)}}{- \B(k)e^{2ik(L-x)}}{0}{
\frac{a(k)}{\alpha(k)}}, \quad \hat J_3 = \curl{
\frac{\overline{\alpha(\bar k)}}{\overline{a(\bar k)}}}{0}{ \lambda
\overline{\B(\bar k)}e^{-2ik(L-x)}}{ \frac{\overline{a(\bar k)}}{
\overline{\alpha(\bar k)}}},$$

$$ J^{(\infty)}_1 = \curl{1}{0}{ \frac{\lambda\overline{B(\bar
k)}e^{2ikx}}{a(k)d(k)}}{1}, \quad J^{(\infty)}_3 = \curl{1}{-
\frac{B(k)e^{-2ikx}}{\overline{a(\bar k)} \overline{d(\bar k)}}}{0}{1}.$$
$$ J^{(\infty)}_2 = \curl{1}{- \frac{b(k)}{\overline{a(\bar
k)}}e^{-2ikx}}{\lambda \frac{\overline{b(\bar k)}}{a(k)}e^{2ikx}}{
\frac{1}{a(k)\overline{a(\bar k)}}}. \eqno (3.24)$$
It can be verified that
$$J_1(x,0,k) = \hat J_1J^{(\infty)}_1,  J_2(x,0,k) = \hat
J_1J^{(\infty)}_2\hat J_3, J_3(x,0,k) = J^{(\infty_)}_3\hat J_3, J_4(x,0,k)
= J^{(\infty)}_3 (J^{(\infty)}_2)^{-1}J^{(\infty)}_1.\eqno (3.25)$$
Let $M^{(1)}(x,t,k)$, $M^{(2)}(x,t,k)$, $M^{(3)}(x,t,k)$, $M^{(4)}(x,t,k)$,
denote $M(x,t,k)$ for $\arg k \in [0,\frac{\pi}{2}],...,\arg k \in
[\frac{3\pi}{2},2\pi]$.  Then the jump condition (2.3.4) becomes

$$ M^{(2)} = M^{(1)}J_1, M^{(2)} = M^{(3)}J_4, M^{(4)} = M^{(1)}J_2,
M^{(4)} = M^{(3)}J_3. \eqno (3.26)$$
Evaluating these equations at $t=0$ and using equations (3.25) we find
$$ M^{(2)}(x,0,k) = (M^{(1)}(x,0,k)\hat J_1)J^{(\infty)}_1, M^{(2)}(x,0,k)
= M^{(3)}(x,0,k) J^{(\infty)}_3(J^{(\infty)}_2)^{-1}J^{(\infty)}_1, $$
$$(M^{(4)}(x,0,k)\hat J^{-1}_3) = (M^{(1)}(x,0,k)\hat J_1)J^{(\infty)}_2,
(M^{(4)}(x,0,k)\hat J^{-1}_3) = M^{(3)}(x,0,k) J^{(\infty)}_3. \eqno
(3.27)$$
Defining $M^{(\infty)}_j(x,k)$, $j=1,...,4$, by
$$M^{(\infty)}_1 = M^{(1)}(x,0,k)\hat J_1(x,k),\quad M^{(\infty)}_2 =
M^{(2)}(x,0,k),
\eqno (3.28a)
$$
$$
M^{(\infty)}_3 = M^{(3)}(x,0,k), \quad M^{(\infty)}_4 = M^{(4)}(x,0,k)\hat
J^{-1}_3(x,k),
\eqno (3.28b)$$
we find that the sectionally holomorphic function $M^{(\infty)}(x,k)$
satisfies the jump conditions
$$ M^{(\infty)}_2 = M^{(\infty)}_1 J^{(\infty)}_1,
M^{(\infty)}_2 = M^{(\infty)}_3J^{(\infty)}_3 (J^{(\infty)}_2)^{-1}
J^{(\infty)}_1,
M^{(\infty)}_4 = M^{(\infty)}_1J^{(\infty)}_2,
M^{(\infty)}_4 = M^{(\infty)}_3J^{(\infty)}_3.$$
These conditions are precisely the jump conditions satisfied by the unique
solution of the RH problem associated with the NLS for $0<x<\infty$,
$0<t<T$ \cite{6}.  Also $\det M^{(\infty)}=1$  and $M^{(\infty)} =
I+O(\frac{1}{k})$, $k\rightarrow \infty$. Moreover, by
a straightforward calculation one can verify that the transformation
(3.28) replaces poles at $\nu_{j}$ by poles at $k_{j}$,
with the residue conditions (2.4.1), (2.4.2), replaced by the
proper residue conditions at $k = k_{j}$ (cf. \cite{6}). Therefore,
$M^{(\infty)}(x,k)$ satisfies the same RH
problem as the RH problem associated with the half-line evaluated at $t=0$.
Hence $q(x,0) = q_0(x)$.

\vskip .2in
\noindent {\bf Proof that $q(0,t) = g_0(t)$, $q_x(0,t) = g_1(t)$}

Let $M^{(0)}(t,k)$ be defined by
$$ M^{(0)}(t,k) = M(0,t,k)G(t,k), \eqno (3.29)$$
where $G$ is given by $G^{(1)},...,G^{(4)}$, for $\arg k \in [0,
\frac{\pi}{2}],...,[\frac{3\pi}{2}, 2\pi]$.  Suppose we can find matrices
$G^{(j)}$ which are holomorphic, tend to $I$ as $k\rightarrow \infty$, and
satisfy
$$ J_1(0,t,k)G^{(2)} = G^{(1)}J^{(0)}, J_2(0,t,k)G^{(4)} = G^{(1)}J^{(0)},
J_3(0,t,k)G^{(4)} = G^{(3)}J^{(0)}, \eqno (3.30)$$
where $J^{(0)}(t,k)$ is defined in Remark 3.2.  Then equations (3.30)
yield $J_4(0,t,k)G^{(2)} = G^{(3)}J^{(0)}$, and equations (3.26), (3.29)
imply that $M^{(0)}(t,k)$ satisfies the RH problem defined in Remark 3.2.
Then Remark 3.2 implies the desired result.

We will show that such $G^{(j)}$ matrices are:
$$G^{(1)} = \curl{ \frac{\alpha(k)}{A(k)}}{c^+(k)e^{4ik^2(T-t)}}{0}{
\frac{A(k)}{\alpha(k)}}, \quad G^{(4)} = \curl{ \frac{\overline{A(\bar
k)}}{\overline{\alpha(\bar k)}}}{0}{\lambda \overline{c^+(\bar
k)}e^{-4ik^2(T -t)}}{ \frac{\overline{\alpha(\bar k)}}{\overline{A(\bar
k)}}}$$

$$ G^{(2)} = \curl{d(k)}{ \frac{-b(k)}{\overline{A(\bar
k)}}e^{-4ik^2t}}{0}{ \frac{1}{d(k)}}, \quad G^{(3)} =
\curl{\frac{1}{\overline{d(\bar k)}}}{0}{\frac{ -\lambda \overline{b( \bar
k)}}{A(k)}e^{4ik^2t}}{\overline{d(\bar k)}}. \eqno (3.31)$$

We recall that in the half line problem the associated
matrices $J^\infty_2(0,t,k)$, $G^{\infty(1)}(t,k)$, $G^{\infty(4)}(t,k)$
satisfy
$$ J^\infty_2(0,t,k)G^{\infty(4)} = G^{\infty(1)}J^{(t)}; \eqno (3.32)$$
for the verification of this equation one uses
$$ a\bar a - \lambda b\bar b =1, A\bar A - \lambda B\bar B =1,  a B-bA
= e^{4ik^2T}c^+ . \eqno (3.33)$$
The matrix $J_2(0,t,k)$ can be obtained from the matrix
$J^{\infty}_2(0,t,k)$ by replacing $a$ and $b$ with $\alpha$ and $\beta$;
furthermore, $\alpha,\beta$, $A,B$ satisfy equations similar to equations
(3.33) where $a$ and $b$ are replaced by $\alpha$ and $\beta$.  Hence,
$G^{(4)}$ and $G^{(1)}$ follow from $G^{\infty(4)}$ and $G^{\infty(1)}$ by
replacing $a$ and $b$ by $\alpha$ and $\beta$; this yields the first two
equations of (3.31).

Having obtained $G^{(1)}$, the first of equations (3.30) yields $G^{(2)}$
(then $G^{(3)}$ follows form symmetry considerations).  Rather than
deriving $G^{(1)}$ we verify that it satisfies the equation
$J_1(0,t,k)G^{(2)} - G^{(1)}J^{(0)}=0$: The (21) and (22) elements are
satisfied identically.  The (11) element is satisfied iff

$$ \delta = \frac{\alpha}{A} + \frac{\lambda \bar B}{A} c^+ e^{4ik^2T};
\eqno (3.34)$$
but
$$ \delta = \frac{\alpha \bar A A}{A} - \lambda \beta\bar B =
\frac{\alpha}{A}(1+\lambda B\bar B) - \lambda \beta \bar B =
\frac{\alpha}{A} + \frac{\lambda \bar B}{A} (\alpha B-\beta A), $$
which equals the rhs of (3.34) in view of the global relation.  The (12)
element is satisfied iff
$$ \frac{\delta b}{d\bar A} + \frac{\B e^{2ikL}}{d} = \frac{\alpha
B}{A\bar A} - \frac{c^+e^{4ik^2T}}{A\bar A}, $$
which using the global relation to replace $c^+\exp (4ik^2T)$,  becomes
$$ \delta b + \bar A\B e^{2ikL} = \beta d; \eqno (3.35)$$
the rhs of this equation is
$$ b(\alpha \bar A - \lambda \beta \bar B) + \bar A\B e^{2ikL} =-\lambda
\beta b \bar B + \bar A(\alpha b + \B e^{2ikL}),$$
which equals the rhs of equation (3.35) using the identity (2.3.14).

Similar to the proof of the equation $q(x,0) = q_{0}(x)$, it can be
verified that the transformation (3.29) replaces the residue conditions
(2.4.1) - (2.4.5) by the residue conditions of Remark 3.2.

\vskip .2in
\noindent {\bf Proof that $q(L,t) = f_0(t)$, $q_x(L,t) = f_1(t)$}

Following arguments similar to the proof above we seek matrices
$F^{(j)}(t,k)$ such  that
$$ J_1(L,t,k)F^{(2)} = F^{(1)}J^{(L)}, J_4(L,t,k)F^{(2)} =
F^{(3)}J^{(L)}, J_3(L,t,k) F^{(4)} = F^{(3)}J^{(L)}. \eqno (3.36)$$
(it is more convenient to use the second of these equations instead of
$J_2(L,t,k)F^{(4)} = F^{(1)}J^{(L)}$, see below).  We will show that such
$F^{(j)}$ matrices are
$$ F^{(1)} = \curl{-1}{0}{\frac{-\lambda \overline{b(\bar
k)}e^{4ik^2t+2ikL}}{\alpha(k)\A(k)}}{-1}, \quad F^{(4)} = \curl{-1}{
\frac{-b(k)e^{-4ik^2t-2ikL}}{\overline{\alpha(\bar k)}\overline{\A (\bar
k)}}}{0}{-1}$$
$$ F^{(3)} = \curl{- \frac{1}{\A(k)}}{
\frac{c^+(k)e^{4ik^2(T-t)-2ikL}}{\overline{d(\bar k)}}}{0}{-\A(k)}, \quad
F^{(2)} = \curl{-\overline{\A(\bar k)}}{0}{ \frac{\lambda
\overline{c^+(\bar k)}e^{-4ik^2(T-t) + 2ikL}}{d(k)}}{- \frac{1}{\overline{ \A
(\bar
k)}}}. \eqno (3.37)$$

The matrix $J_4(L,t,k)$ can be written in the form
$$J_4(L,t,k) = \overline{\Lambda(\bar k)}\tilde J_4(t,k)\Lambda (k), \eqno
(3.38)$$
where
$$ \Lambda (k) = \diag \left( \frac{e^{2ikL}}{d(k)}, e^{-2ikL}d(k)\right),
\eqno (3.39)$$
$$\tilde J_4(t,k) = \curl{1}{- \frac{\tilde
\beta(k)}{(-d(k)e^{-2ikL})}e^{-4ik^2t}}{\frac{\lambda \overline{\tilde
\beta (\bar k)}e^{4ik^2t}}{(-\overline{d(\bar k)}e^{2ikL})}}{
\frac{1}{d(k)\overline{d(\bar k)}}}, \eqno (3.40)$$
$$ \tilde \beta (k) = A(k)b(k) - B(k)a(k). \eqno (3.41)$$
Thus the second of equations (3.36) becomes
$$\tilde J_4(t,k) \Lambda (k)F^{(2)} = (\overline{\Lambda(\bar
k)})^{-1}F^{(3)}J^{(L)}. \eqno (3.42)$$
The functions $\tilde \beta(k)$, $\tilde \alpha (k)
=-e^{2ikL}\overline{d(\bar k)}$, $\A(k)$, $\B(k)$ satisfy the following
equations
$$ \tilde \alpha \bar{\tilde \alpha} - \lambda \tilde \beta \bar{\tilde
\beta} = 1, \A \bar \A - \lambda \B \bar \B =1, \tilde \alpha \B - \tilde
\beta \A = e^{4ik^2T} c^+(k); \eqno (3.43)
$$
indeed the first of these equations is $\det \tilde J_4=1$ and the third of
these equations is the global relation.  Thus, comparing equation (3.42)
with (3.32) it follows that $(\overline{\Lambda(\bar k)})^{-1}F^{(3)}$ can
be obtained from $G^{(\infty)(1)}$ with $a$, $b$, $A$, $B$ replaced by
$\tilde \alpha$, $\tilde \beta$, $\A$, $\B$; this yields the second two of
equations (3.37).

Having determined $F^{(2)}$, the first of equation (3.36) yields $F^{(1)}$.
Rather than deriving $F^{(1)}$ we show that the equation
$F^{(1)}J^{(L)}-J_1(L,t,k)F^{(2)}=0$ is valid: The (12) and (22) elements are
satisfied identically.  The (21) element is satisfied iff
$$ \frac{\bar \A\bar Be^{2ikL}}{d\alpha} - \frac{a \bar
c^+e^{-4ik^2T+2ikL}}{d\alpha} = \frac{\bar be^{2ikL}}{\alpha \A} +
\frac{\bar \B}{\A}.$$
Using the global relation to replace $\exp[-4ik^2T]$, and then using
$1-a\bar a = -\lambda b\bar b$, the above equation  becomes
$$\A (a\bar \B + \bar b \bar \A e^{2ikL}) = \bar be^{2ikL} + \bar \B
\alpha;$$
using the definition of $\alpha$, as well as $\A \bar \A  - \lambda \B \bar
\B =1$, the above equation becomes an identity.  The direct verification
of the (11) element can be avoided by appealing to the equality of the
determinants.

Similar to the previous case, the transformation
$$
M(L,t,k) \mapsto M^{(L)}(t,k) = M(L,t,k)F(t,k)
$$
maps the Riemann-Hilbert problem of Theorem 3.1 to the
Riemann-Hilbert problem of Remark 3.3.\hfill {\bf QED}

\section{The Analysis of the Global Relation}

Evaluating equation (2.1.8) at $x=0$, instead of
$x=0$, $t=T$, we find instead of equation (1.4) the
following equation:
$$
(a(k)\A(t,k) + \overline{\lambda b(\bar{k})}e^{2ikL}
\B(t,k))B(t,k) -
(b(k)\A(t,k) + \overline{\lambda a(\bar{k})}e^{2ikL}
\B(t,k))A(t,k) \eqno (4.1)
$$
$$
= e^{4ik^2t}c^{+}(k,t), \quad k \in {\mathbb C},
$$
where $c^{+}(k,t)$ has the form
$$
c^{+}(k,t) = \int_{0}^{L}e^{2ikx}c(k,x,t)dx, \eqno (4.2)
$$
and $c(k,x,t)$ is entire function in $k$ which,
together with its $x$ and $t$ derivatives, is
of $O(1)$ as $k \to \infty$ and $k$  is in the first quadrant.
This in turn implies that $c^+(k,t)$ is an entire function in $k$ which is
of $O(1/k)$ as $k
\rightarrow
\infty$, Im$k >0$; in fact,
$$
c^{+}(k) = O\left(\frac{1+ e^{2ikL}}{k}\right), \quad k \to \infty.
$$

For the analysis of equation (4.1) we will use the
following two identities:
$$
\int_{\partial{D_{1}}}k\left[\int_{0}^{t}
e^{4ik^2(\tau - t')}K(\tau,t)d\tau \right]dk = \frac{\pi}{4}K(t',t),
\eqno (4.3)
$$

$$
\int_{\partial{D^{0}_{1}}}\frac{k^2}{\Delta(k)}\left[\int_{0}^{t}
e^{4ik^2(\tau - t')}K(\tau,t)d\tau \right]dk
$$
$$
= \int_{\partial{D^{0}_{1}}}\frac{k^2}{\Delta(k)}\left[\int_{0}^{t'}
e^{4ik^2(\tau - t')}K(\tau,t)d\tau - \frac{K(t',t)}{4ik^2} \right]dk,
\eqno (4.4)
$$
where
$$
t >0, \quad t' >0, \quad t' < t,
$$
$\partial{D_{1}}$ denotes the union of the contour $(i\infty, 0]$
and of the contour $[0, \infty)$ (i.e. the oriented boundary of the
first quadrant), $\partial{D^{0}_{1}}$ denotes the contour obtained by
deforming $\partial{D_{1}}$ to the contour passing above the points
$ k = \frac{\pi m}{2L}$, $ n \in {\mathbb Z}^{+}$, $K(\tau, t)$ is
a smooth function of the arguments indicated, and
$$
\Delta(k) = e^{2ikL} - e^{-2ikL}.
\eqno(4.5)
$$
Indeed, in order to derive equation (4.4) we rewrite the lhs of this
equation as the rhs plus the term
$$
\int_{\partial{D^{0}_{1}}}\frac{k^2}{\Delta(k)}\left[\int_{t'}^{t}
e^{4ik^2(\tau - t')}K(\tau,t)d\tau + \frac{K(t',t)}{4ik^2} \right]dk.
$$
The integrand of the above integral is analytic and bounded in the
domain of the complex-$k$ plane enclosed by $\partial{D^{0}_{1}}$.
Also its zero order term (with respect to $(k^2)^{-1}$) contains
the oscillatory factor $e^{4ik^2(t - t')}$, thus Jordan's lemma
implies that this term vanishes.  Similarly,  in order to derive
equation (4.3) we rewrite the lhs of this equation in the form
$$
\int_{\partial{D_{1}}}k\left[\int_{0}^{t'}
e^{4ik^2(\tau - t')}K(\tau,t)d\tau \right]dk +
\int_{\partial{D_{1}}}k\left[\int_{t'}^{t}
e^{4ik^2(\tau - t')}K(\tau,t)d\tau \right]dk.
\eqno (4.6)
$$
The contour $\partial{D_{1}}$ involves the contour $[0, \infty)$,
which can be mapped to the contour $[0, -\infty)$ by replacing
$k$ with $-k$, thus $\partial{D_{1}}$ can be replaced by
$\partial{D_{2}}$ which denotes the union of the contours $(i\infty, 0]$
and $[0, -\infty)$. Hence replacing  $\partial{D_{1}}$ by $\partial{D_{2}}$
in the first integral of the expression (4.6) we have
$$
\int_{\partial{\hat{D}_{2}}}\left[\int_{0}^{t'}
e^{4ik^2(\tau - t')}kK(\tau,t)d\tau - \frac{K(t',t)}{4ik} \right]dk
+ \frac{K(t',t)}{4i}\int_{\partial{\hat{D}_{2}}}\frac{dk}{k}
$$
$$
+ \int_{\partial{\hat{D}_{1}}}\left[\int_{t'}^{t}
e^{4ik^2(\tau - t')}kK(\tau,t)d\tau + \frac{K(t',t)}{4ik} \right]dk
- \frac{K(t',t)}{4i}\int_{\partial{\hat{D}_{1}}}\frac{dk}{k},
$$
where $\hat{D}$ indicates that we have indented the contour $D$
to avoid $k =0$. The first  and the third integral vanish, since the
first and the third integrands are analytic and decaying in
$\hat{D}_{2}$ and $\hat{D}_{1}$ respectively, and their
 first order terms (with respect to $k^{-1}$) contain
the oscillatory factors $e^{-4ik^2t'}$ and $e^{4ik^2(t - t')}$
respectively.
The remaining two integrals equal
$$
\frac{K(t',t)}{4i}\int_{0}^{\pi}id\theta = \frac{\pi}{4}K(t',t).
$$

We will now show that the global relation (4.1) can be
explicitely solved for $f_{1}(t)$ and $g_{1}(t)$. To avoid
routine technical complications we shall consider here
the case of zero initial conditions, which yields $a(k)\equiv 1$
and $b(k) \equiv 0 $. We will show that in this case the
expressions for $f_{1}(t)$ and $g_{1}(t)$ are:
$$
\frac{i\pi}{4}f_{1}(t) =
\int_{\partial{D^{0}_{1}}}\frac{2k^2}{\Delta(k)}\left[
\hat M_{1}(t,k) - \frac{g_{0}(t)}{2ik^2}\right]dk
- \int_{\partial{D^{0}_{1}}}k^{2}\frac{\Sigma(k)}{\Delta(k)}\left[
\hat{\MM}_1(t,k) - \frac{f_{0}(t)}{2ik^2}\right]dk
$$
$$
+ \int_{\partial{D^{0}_{1}}}\frac{k}{\Delta(k)}\left[
F(t,k) - F(t,-k)\right]dk,\eqno(4.7)
$$

$$
-\frac{i\pi}{4}g_{1}(t) =
\int_{\partial{D^{0}_{1}}}\frac{2k^2}{\Delta(k)}\left[
\hat{\MM}_{1}(t,k) - \frac{f_{0}(t)}{2ik^2}\right]dk
- \int_{\partial{D^{0}_{1}}}k^{2}\frac{\Sigma(k)}{\Delta(k)}\left[
\hat{M}_1(t,k) - \frac{g_{0}(t)}{2ik^2}\right]dk
$$
$$
- \int_{\partial{D^{0}_{1}}}\frac{k}{\Delta(k)}\left[
e^{-2ikL}F(t,k) - e^{2ikL}F(t,-k)\right]dk,\eqno(4.8)
$$
where
$$
\Sigma(k) = e^{2ikL} + e^{-2ikL}, \eqno(4.9)
$$
and
$$
F(t,k) = \frac{if_{0}(t)}{2}e^{2ikL}\hat{\MM}_{2}
-\frac{ig_{0}(t)}{2}\hat{M}_{2}
+ \left[\overline{\hat{\LL}_2}
-i\lambda \frac{f_{0}(t)}{2}\overline{\hat{\MM}_1}
+k\overline{\hat{\MM}_2}\right]
\left[\hat{L}_1
-i\frac{g_{0}(t)}{2}\hat{M}_2
+k\hat{M}_1\right]
$$
$$
-e^{2ikL}\left[\overline{\hat{L}_2}
-i\lambda \frac{g_{0}(t)}{2}\overline{\hat{M}_1}
+k\overline{\hat{M}_2}\right]
\left[\hat{\LL}_1
-i\frac{f_{0}(t)}{2}\hat{\MM}_2
+k\hat{\MM}_1\right]. \eqno(4.10)
$$
Indeed, substituting in equation (4.1) (with $a \equiv 1$ and $b \equiv
0$) the expressions for $A$, $B$ from equations (1.12) as well as the
analogous expressions for  $\A$, $\B$ we find
$$
-2\int^t_0 e^{4ik^2\tau}L_1(t,2\tau-t)d\tau
+ 2e^{2ikL}\int^t_0 e^{4ik^2\tau}{\LL}_1(t,2\tau-t)d\tau
$$
$$
=2k\int^t_0 e^{4ik^2\tau}M_1(t,2\tau-t)d\tau
- 2ke^{2ikL}\int^t_0 e^{4ik^2\tau}{\MM}_1(t,2\tau-t)d\tau
$$
$$
+ e^{4ik^{2}t}F(t,k) + e^{4ik^{2}t}c^{+}(t,k). \eqno (4.11)
$$
Regarding $F(t,k)$ we note that we first write $F(t,k)$
in terms of $\left\{  L_j(t,k), \,  M_j(t,k),
\, {\LL}_j(t,k), \, {\MM}_j(t,k) \right\}^2_1$
and then use equations (1.15) to rewrite $F(t,k)$ in the
form (4.10). Using integration by parts it
follows that $e^{4ik^{2}t}F(t,k) = O(1/k^{2})$ as $k \to \infty$.

Replacing $k$ by $-k$ in equation (4.11) and solving the resulting
equation as well as equation (4.11) for the two integrals appearing in the
l.h.s. of equation (4.11), we find the following:
$$
2\int^t_0 e^{4ik^2\tau}{\LL}_1(t,2\tau-t)d\tau
= \frac{4k}{\Delta(k)}\int^t_0 e^{4ik^2\tau}M_1(t,2\tau-t)d\tau
$$
$$
- 2k\frac{\Sigma(k)}{\Delta(k)}\int^t_0 e^{4ik^2\tau}{\MM}_1(t,2\tau-t)d\tau
+ \frac{G(t,k) - G(t,-k)}{\Delta(k)},
\eqno(4.12)
$$
and
$$
-2\int^t_0 e^{4ik^2\tau}{L}_1(t,2\tau-t)d\tau
= \frac{4k}{\Delta(k)}\int^t_0 e^{4ik^2\tau}{\MM}_1(t,2\tau-t)d\tau
$$
$$
- 2k\frac{\Sigma(k)}{\Delta(k)}\int^t_0 e^{4ik^2\tau}{M}_1(t,2\tau-t)d\tau
-\frac{e^{-2ikL}G(t,k) - e^{2ikL}G(t,-k)}{\Delta(k)},
\eqno(4.13)
$$
where
$$
G(t,k) = e^{4ik^{2}t}F(t,k) + e^{4ik^{2}t}c^{+}(t,k).
$$
We multiply equation (4.12) by $k\exp(-4ik^2t')$, $t'>0$, $t' < t$,
and integrate over $\partial{D}_{1}^{0}$. In this respect we note that
the function
$$
k\left[\frac{c^{+}(t,k) - c^{+}(t,-k)}{\Delta(k)}\right],
$$
is analytic and bounded in the interior of $\partial{D}_{1}^{0}$,
thus the integral of the term,
$$
ke^{4ik^2(t-t')}\left[\frac{c^{+}(t,k) - c^{+}(t,-k)}{\Delta(k)}\right],
$$
vanishes. The integrals involving ${\LL}_{1}$ and $M_{1}$ can be
computed using equations (4.3) and (4.4), respectively. Also, the
term involving ${\MM}_{1}$ can be computed using equation (4.4)
with $1/{\Delta}$ replaced by $\Sigma/{\Delta}$. In this way we find that
$$
\frac{\pi}{2}{\LL}_1(t,2t'-t)
= \int_{\partial{D}_{1}^{0}}\frac{4k^2}{\Delta(k)}\left[\int^{t'}_0
e^{4ik^2(\tau - t')}M_1(t,2\tau-t)d\tau-
\frac{M_{1}(t,2t'-t)}{4ik^2}\right]dk
$$
$$
-
\int_{\partial{D}_{1}^{0}}2k^{2}\frac{\Sigma(k)}{\Delta(k)}\left[\int^{t'}_0
e^{4ik^2(\tau - t')}{\MM}_1(t,2\tau-t)d\tau-
\frac{{\MM}_{1}(t,2t'-t)}{4ik^2}\right]dk
$$
$$
+\int_{\partial{D}_{1}^{0}}\frac{k}{\Delta(k)}
e^{4ik^{2}(t-t')}(F(t,k) - F(t,-k))dk.
$$
Evaluating this equation at $t=t'$ and using the first and third
equations of equations (1.10) and (1.11) respectively we find equation
(4.7). The derivation of equation (4.8) is similar.

\section{Conclusions}
We have analysed the Dirichlet problem for the nonlinear Schr\"odinger equation on the finite interval, see equations
(1.1).  In particular:

(i) Given the Dirichlet data $q(0,t) = g_0(t)$ and $q(L,t) = f_0(t)$, we have characterised the Neumann
boundary values $q_x(0,t) = g_1(t)$ and $q_x(L,t) = f_1(t)$ through a system of nonlinear
ODEs for the functions (1.14).  The functions $\{ \hat L_j, \hat M_j\}^2_{j=1}$ satisfy equations
(1.16), the functions $\{ \hat \LL_j, \hat \MM_j\}^2_{j=1}$ satisfy similar equations, and the Neumann
boundary values are given by equations (4.7) and (4.8).

(ii) Given the initial condition $q(x,0) = q_0(x)$ we have defined $\{ a(k),b(k)\}$, see definition 3.1.
Given $g_0(t)$ and $g_1(t)$ we have defined $\{ A(k),B(k)\}$, and given $f_0(t)$ and $f_1(t)$ we have defined $\{
\A(k),\B(k)\}$, see definitions 3.2 and 3.3.

(iii) Given $\{ a(k), b(k), A(k), B(k), \A(k), \B(k)\}$ we have defined a Riemann-Hilbert problem for $M(x,t,k)$ and then
we have defined $q(x,t)$ in terms of $M$.  We have shown that $q(x,t)$ solves the nonlinear Schr\"odinger, and that
$$q(x,0) = q_0(x), q(0,t) = g_0(t), q_x(0,t) = g_1(t), q(L,t) = f_0(t), q_x(L,t) = f_1(t),$$
see theorem 3.1.

A general method for analyzing initial-boundary
value problems
for integrable PDEs was introduced in \cite{5}.  This method is based on the {\it
simultaneous} spectral analysis of the two eigenvalue equations forming
the associated
Lax pair, and on the investigation of the global relation satisfied by the
relevant
spectral functions.  The rigorous implementation of this method to the NLS
on the half
line was presented in \cite{6}.  Analogous results for the sine Gordon, KdV
(with dominant
surface tension) and modified KdV equation, were presented in \cite{7} and \cite{8}.
 The most
difficult step of this method is the analysis of the global relation.
Rigorous results
for this problem were obtained in \cite{6} by analyzing the global
relation, which is a scalar equation relating $g_0$, $g_1$, $\Phi$,
together with the equation satisfied by $\Phi$, which is a vector
equation relating $g_0$, $g_1$, $\Phi$.  This analysis is quite
complicated and this is partly
due to the fact that these two equations are {\it coupled}.  An important
development
in this direction was announced in \cite{4}, where it was shown that if one
uses the
Gelfand-Levitan-Marchenko representation for $\Phi$, then the above two
equations can
be {\it decoupled}.  Indeed, the global relation can be {\it explicitly}
solved for
$g_1$ in terms of $g_0$ and $\Phi$ (or more precisely in terms of $\hat
L_j$, $\hat M_j$, $j=1,2$).
In this paper we have extended the results of \cite{4} and \cite{6} to the case
that the NLS is defined on a finite interval instead of the
half-line.

The analysis of the analogous problem for the modified KdV equation
but without using the new results of \cite{4} is presented in
\cite{8}, \cite{9}; see also \cite{10}.

A different approach to this problem, which instead of the
Riemann-Hilbert formalism uses the periodic extension to the whole
line, is presented in \cite{12}.

For integrable evolution PDEs on the half-line,
there exist
particular boundary conditions for which the nonlinear Volterra integral
equations can
be avoided.  For these boundary conditions, which we call {\it
linearizable}, the
global relation yields directly $S(k)$ in terms of $s(k)$ and the
prescribed boundary
conditions \cite{6}, \cite{7}.  Different aspects of linearizable boundary
conditions have been
studied by a number of authors, see for example \cite{10}--\cite{13}.  The analysis
of linearizable
boundary conditions for the NLS on a finite domain will be presented
elsewhere.  Here
we only note that {\it $x$-periodic} boundary conditions belong to the
linearizable
class.  In this case $S(k) = S_L(k)$ and the global relation simplifies.  The
analysis of this simplified global relation, together with the results
presented in
this paper, yield a {\it new} formalism for the solution of this classical
problem.

\section*{Acknowledgements}

This work was partially supported by the EPSRC. The second author was
also supported in part by the NSF Grant DMS-0099812.

\end{document}